\pdfoutput=1

\documentclass{iaus}
\usepackage{graphicx}
\def\apj{ApJ}                 % Astrophysical Journal
\def\apjl{ApJ}                % Astrophysical Journal, Letters
\def\aap{A\&A}                % Astronomy and Astrophysics
\def\mnras{MNRAS}             % Monthly Notices of the RAS
\newcommand{\mearth}{M_\oplus}

\newcommand{\beq}{\begin{equation}}
\newcommand{\eeq}{\end{equation}}
\newcommand{\beqa}{\begin{eqnarray}}
\newcommand{\eeqa}{\end{eqnarray}}

\title[Planetary population synthesis: type I migration] {Application of recent results on the orbital migration of low mass planets:\\ convergence zones}
\author[C. Mordasini, K.-M. Dittkrist, Y. Alibert, H. Klahr, W. Benz, T. Henning]   %% give here short author list %%
{C. Mordasini$^1$, K.-M. Dittkrist$^1$, Y. Alibert$^2$, H. Klahr$^1$, W. Benz$^2$ \and T. Henning$^1$}

\affiliation{$^1$Max Planck Institute for Astronomy, K¬\"onigstuhl 17, D-69117 Heidelberg, Germany \\ email: {\tt mordasini@mpia.de} \\[\affilskip]
$^2$Physikalisches Institut, Sidlerstrasse 5, CH-3012 Bern, Switzerland}

\pubyear{2010}
\volume{276}  %% insert here IAU Symposium No.
\pagerange{1--4}
\jname{The Astrophysics of Planetary Systems} %: Formation, Structure, and Dynamical Evolution
\editors{A. Sozzetti, M. G. Lattanzi \& A. P. Boss, eds.}
\begin{document}

\maketitle

\begin{abstract}
Previous models of the combined growth and migration of protoplanets needed large ad hoc reduction factors for the type I migration rate as found in the isothermal approximation. In order to eliminate these factors, a simple semi-analytical model is presented that incorporates recent results on the migration of low mass planets in non-isothermal disks. It allows for outward migration. The model is used to conduct planetary populations synthesis calculations. Two points  with zero torque are found in the disks. Planets migrate both in- and outward towards these convergence zones. They could be important for accelerating planetary growth by concentrating matter in one point.  We also find that the updated type I migration models allow the formation of both close-in low mass planets, but also of giant planets at large semimajor axes. The problem of too rapid migration is significantly mitigated. 

%% add here a maximum of 10 keywords, to be taken form the file <Keywords.txt>
\end{abstract}

\firstsection % if your document starts with a section,
              % remove some space above using this command.
\section{Introduction}
The timescales of orbital migration of low mass  planets  as found  in linear, isothermal type I migration models are very short for typical protoplanetary disk conditions. They are in particular shorter than typical growth timescales \cite[(Tanaka et al. 2002)]{tanakaetal2002}. This means that most protoplanets would fall into the star before reaching a mass  allowing giant planet formation ($\sim 10\, \mearth$). It is therefore not surprising that previous planetary population synthesis models  have found that isothermal type I rates need to be reduced by large factors (10-1000) in order to reproduce the observed semimajor axis distribution of extrasolar planets (\cite[Ida \& Lin 2008]{idalin2008}, \cite[Mordasini et al. 2009b]{mordasinietal2009b}, \cite[Schlaufman et al. 2009]{schlaufmanetal2009}). 

The necessity of such arbitrary reduction factors indicates a significant shortcoming in the understanding of the migration process. Additionally one finds that with universal reduction factors (independent of planet mass and distance), it is impossible to reproduce at the same time giant planets at several AU and close-in low mass planets  \cite[(Mordasini et al. 2009b)]{mordasinietal2009b}. This is inconsistent with observations \cite[(Howard et al. 2010)]{howardetal2010}.  

%\section{Non-isothermal type I migration}
Several mechanisms were proposed that could slow down type I migration. An approach that recently gained significant attention was a more realistic description of the thermodynamics in the protoplanetary disk, in order to drop the simplification of isothermality (e.g. \cite[Paardekooper \& Mellema 2006]{paardekoopermellema2006}, \cite[Kley \& Crida 2008]{kleycrida2008},  \cite[Kley et al. 2009]{kleyetal2009}, \cite[Paardekooper et al. 2010]{paardekooperetal2010}, \cite[Baruteau \& Lin 2010]{baruteaulin2010}). 

It was found that the migration rates (and even the direction of migration) in such more realistic models can be very different from the ones found in the isothermal limit. There are strong dependences on disk properties, like the temperature and the gas surface density gradient or the opacity, leading to different sub-regimes of type I migration. 

\section{Updated type I migration model}
We present results of incorporating these improved type I migration rates into the planet formation code of \cite[Alibert et al. (2005)]{alibertetal2005}. The new  migration model, which will be presented in details in \cite[Dittkrist et al. (in prep.)]{dittkristetalinprep}, includes the following mechanisms:

\begin{figure}
     \begin{minipage}{0.5\textwidth}
	      \centering
       \includegraphics[width=\textwidth]{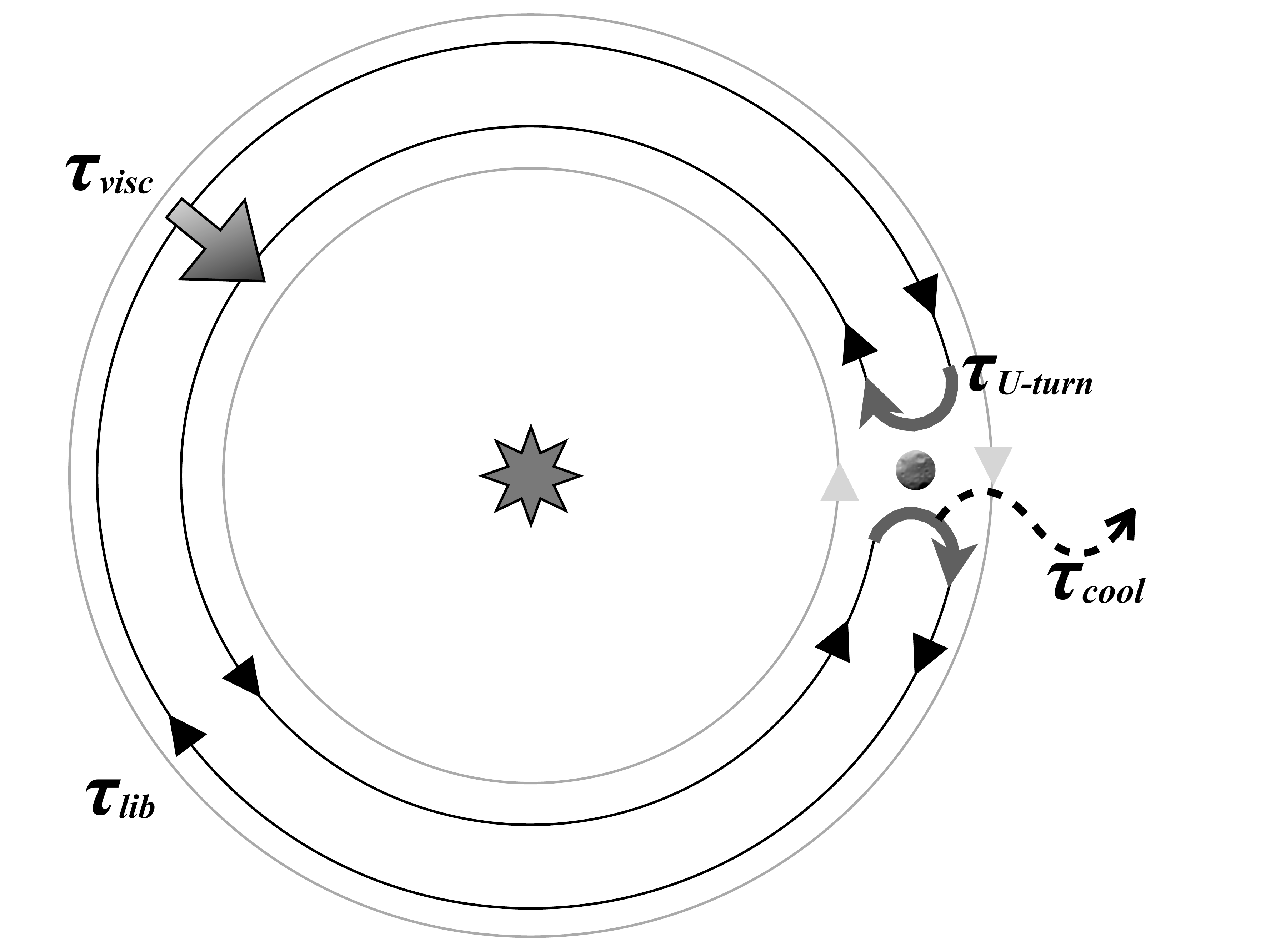}
     \end{minipage}\hfill
     \begin{minipage}{0.5\textwidth}
      \centering
       \includegraphics[width=\textwidth]{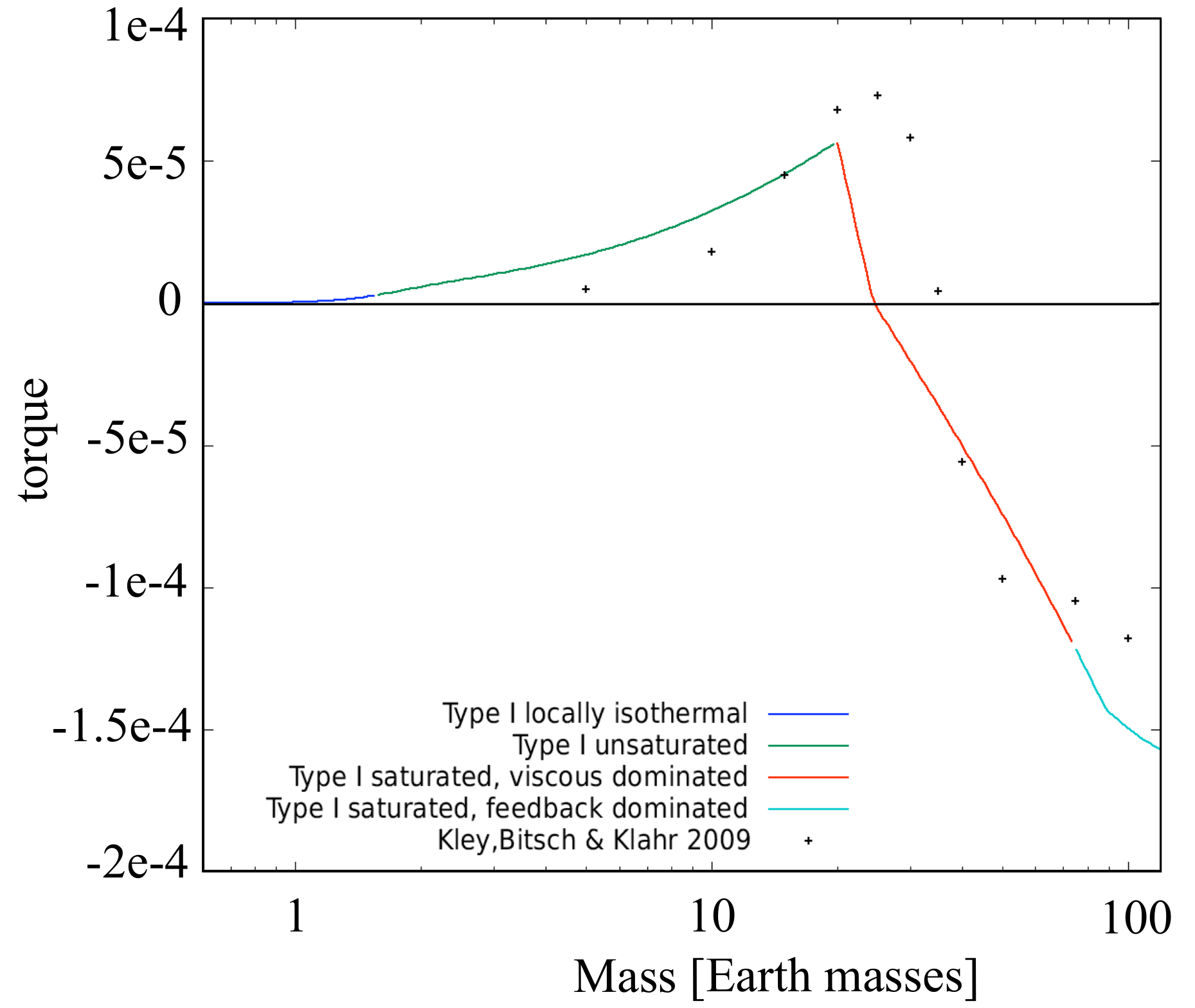}
     \end{minipage}
    \caption{Left panel: Schematic representation of the  timescales relevant to the migration regime of a low mass planet. Trajectories of different gas parcels are shown in a frame of reference rotating with the planet (big black circle) around the star (center). Right panel: Specific torque as a function of planetary mass from the analytical model (lines, indicating different regimes), compared to the radiation-hydrodynamic simulations of \cite[Kley et al. (2009)]{kleyetal2009} (crosses). }\label{mordasinifig:vistnt} 
 \end{figure}
 
{\it Isothermal vs. adiabatic regime}
Gas parcels in the horseshoe region make a sharp U-turn close to the planet (Fig.\,\ref{mordasinifig:vistnt}, left). If during such a U-turn, the gas can cool quickly enough by radiation to equilibrate with the surrounding gas, the gas behaves in a locally isothermal way. On the other hand, if the U-turn timescale $\tau_{\rm U-turn}$ is short compared to the cooling timescale $\tau_{\rm cool}$, then the gas parcel keeps its entropy, and the process is approximately adiabatic. These two situations lead to different density distributions around the planet, which eventually translate into different torques. For both regimes we use the results of \cite[Paardekooper et al. (2010)]{paardekooperetal2010} for the migration rate.

{\it Saturated vs. unsaturated regime}  
The horseshoe region only contains a finite reservoir of angular momentum, which can cause the torque originating from this region to disappear after a finite time. In order to check if such a saturation of the horseshoe drag occurs, we compare the libration timescale $\tau_{\rm lib}$ and the viscous timescale $\tau_{\rm visc}$ across the corotation region (Fig.\,\ref{mordasinifig:vistnt}, left). If  $\tau_{\rm visc}<\tau_{\rm lib}$ sustained outward migration is possible. In the other case, Lindblad torques (plus some residual horseshoe drag) are acting, driving usually inward migration. In some cases, the (outward) migration of the planet itself can keep the corotation region auto-unsaturated in a feedback effect. 

{\it Reduction of the gas surface density} 
As the width of the horseshoe region increases with the planetary mass \cite[(e.g. Masset et al. 2006)]{massetetal2006}  and the viscous timescale  increase with this width, saturation sets in when the planet reaches some mass, of order $10\ \mearth$. Such planets are  particularly vulnerable to a rapid migration into the star, as the migration rate increases with planetary mass. On the other hand, starting gap formation reduces the gas surface density around the planet, reducing the torques. This is taken into account using the results of \cite[Crida \& Morbidelli (2007)]{cridamorbidelli2007}. 

{\it Gap opening criterion} The transition to type II migration is built on the results of \cite[Crida et al. (2006)]{cridaetal2006}. 

The right panel of Fig.\,\ref{mordasinifig:vistnt} shows the  specific torque (torque per mass unit) as a function of planetary mass. Positive torques correspond to outward migration.  The line represents the result from our semi-analytical model, while crosses represent the 3D radiation-hydrodynamic simulations of \cite[Kley et al. (2009)]{kleyetal2009}, indicating good agreement.  

\section{Convergence zones and migration tracks}
\begin{figure}
     \begin{minipage}{0.48\textwidth}
	      \centering
       \includegraphics[width=\textwidth]{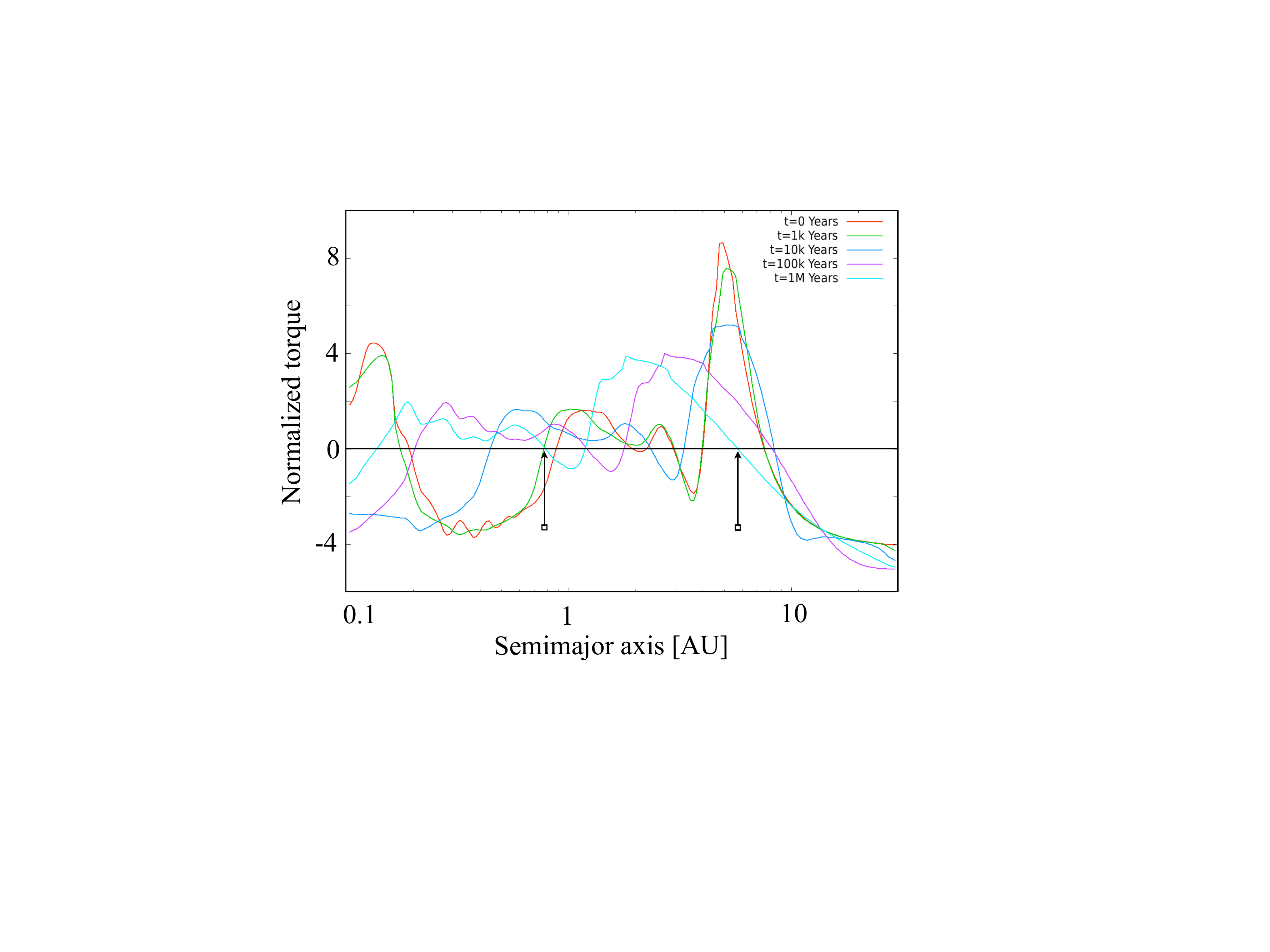}
     \end{minipage}\hfill
     \begin{minipage}{0.5\textwidth}
      \centering
       \includegraphics[width=\textwidth]{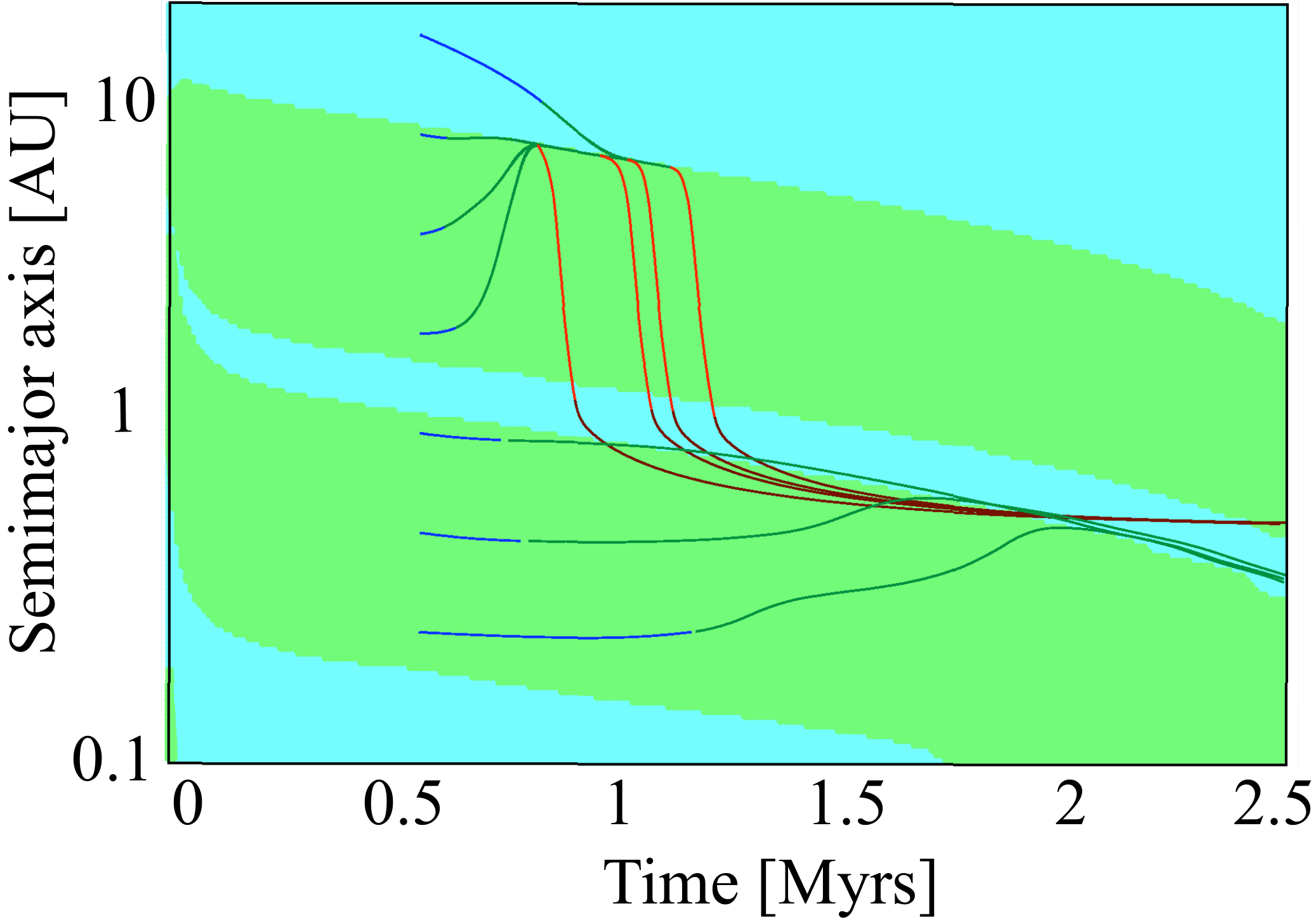}
     \end{minipage}
    \caption{Left panel: Normalized torque as a function of semimajor axis at five times of the disk evolution. The vertical arrows indicate the locations of the convergence zones at 1 Myr. Right panel: The background shows the regions of inward (blue) and outward (green) migration in the disk. The lines show the migration tracks of seven planetary embryo inserted at different starting locations. Colors indicate migration regimes. Blue: locally isothermal. Green: adiabatic, unsaturated. Red: adiabatic, saturated. Brown: type II.}\label{mordasinifig:ta} 
 \end{figure}

With this model at hand, we have studied the migration of growing planets embedded in 1+1D standard alpha disk models (see \cite[Lyra et al. 2010)]{lyraetal2010}. 

The left panel of Fig.\,\ref{mordasinifig:ta} shows the normalized torque in the adiabatic, unsaturated regime as a function of semimajor axis and time for a typical disk. The torque depends on the gradients of the temperature and gas surface density which change because of opacity transitions. There are special zones  where the torque vanishes, and where the torque gradient is negative. This means that inward of these points, migration is directed outward, and outward of them, it is directed inward, so that these locations are migration traps onto which migrating planets converge.

The right panel shows the direction of migration as a function of time and semimajor axis. Green corresponds to outward, blue to inward migration. All protoplanets in the convergence zones migrate towards the stable zero torque locations. We find two convergence zones in agreement with \cite[Lyra et al. (2010)]{lyraetal2010}, which seem to be a generic property. We note that the convergence zones are several AU wide, and therefore could concentrate a lot of matter in one point. This could have important implication for planetary growth, a process very recently studied by \cite[Sandor et al. (2011)]{sandoretal2011}. We also see that  the convergence  zones themselves move inward on a viscous timescale \cite[(Paardekooper et al. 2010)]{paaredekooperetal2010}. 

The lines show migration tracks of seven embryos (initial mass $0.6\,\mearth$) migrating and growing in the disk. Each planet was simulated separately. The inner three protoplanets migrate outward to the inner convergence zone and remain attached to it for the rest of the evolution. They remain low mass planets (3 to 6 $\mearth$).  The outer  four protoplanets migrate both inward and outward to the outer convergence zone. They stay there until the corotation torque saturates because of the mass growth and fast inward migration sets in.  Around 1 AU however, the planetary cores become so massive that gas runaway accretion is triggered, and the migration changes into the slow (planet dominated) type II regime. The final masses of these planets are between 4 to 6 Jupiter masses.

\section{Planetary population synthesis}
\begin{figure}
     \begin{minipage}{0.48\textwidth}
	      \centering
       \includegraphics[width=\textwidth]{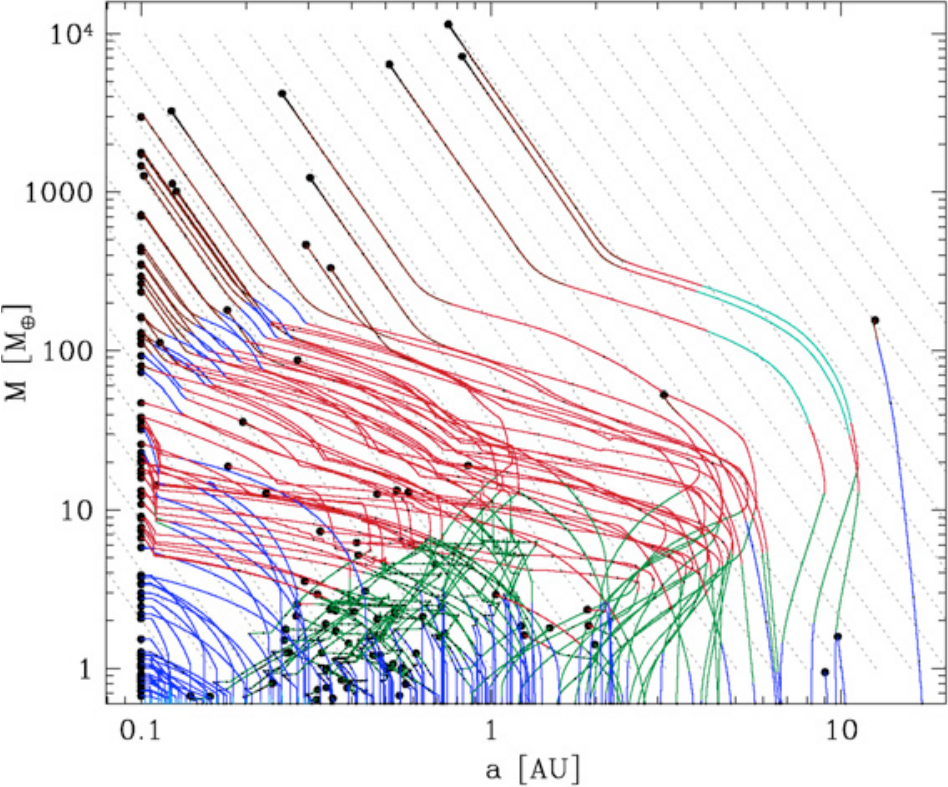}
     \end{minipage}\hfill
     \begin{minipage}{0.48\textwidth}
      \centering
       \includegraphics[width=\textwidth]{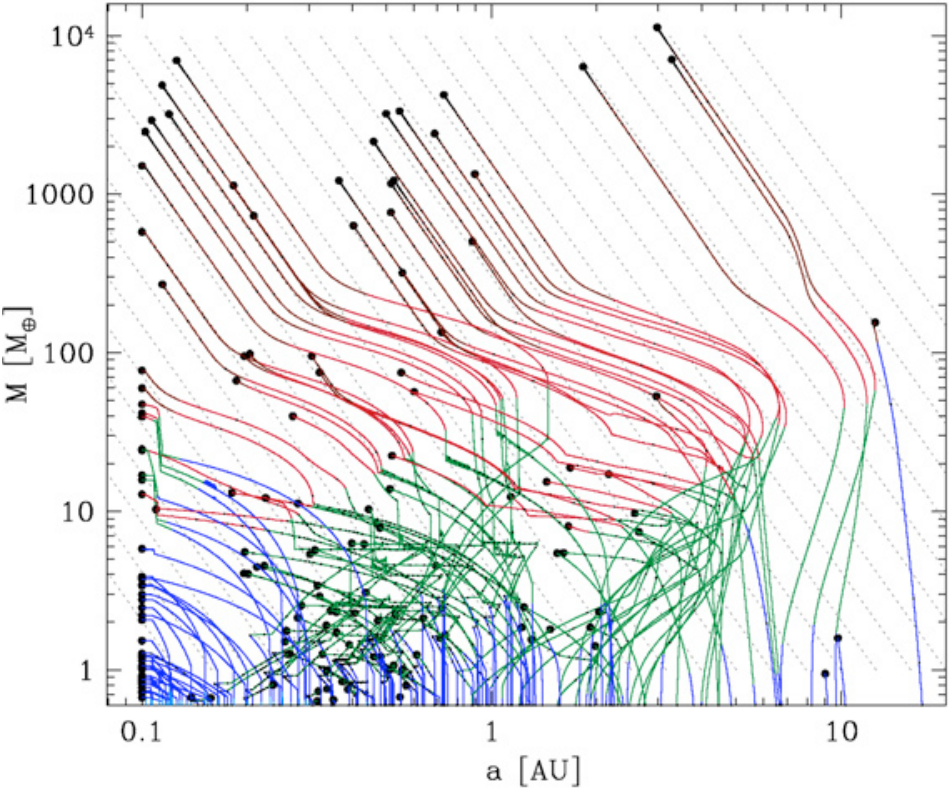}
     \end{minipage}
    \caption{Concurrent growth and migration of protoplanets in the mass-distance plane. Planetary embryos start with an initial mass of $0.6\ \mearth$ at different initial semimajor axes. Final positions are indicated with large black dots. Colors indicate different migration regimes as in Fig. \ref{mordasinifig:ta}. The left panel shows the nominal model, while for the simulation on the right, saturation is assumed to set in at a four times larger mass.    }\label{mordasinifig:pps} 
 \end{figure}
 
Figure \ref{mordasinifig:pps} shows planetary formation tracks in the mass-distance plane, illustrating how planetary embryos concurrently grow and migrate using the undated type I migration model. Other settings and probability distributions are similar as in \cite[Mordasini et al. (2009a)]{mordasinietal2009a}.  Planets were arbitrarily stopped when they migrate to 0.1 AU.

One sees that planets starting inside the inner convergence zone quickly migrate inward leading to close-in low mass planets. Further out, outward migration frequently occurs.  Especially in the right panel, the imprint of the two convergence zones can be seen by two groups of giant planets. In the nominal model in the left panel, many embryos still migrate to 0.1 AU, but much less than with the full rate of \cite[Tanaka et al. (2002)]{tanakaetal2002}. It is clear that (giant) planet formation is no more suppressed by the loss of the embryos into the star. These results will be discussed in details in \cite[Dittkrist et al. (in prep.)]{dittkristetalinprep}.

\end{document}